\documentclass[a4paper]{article}
\makeatletter
\def\@seccntformat#1{\@ifundefined{#1@cntformat}%
   {\csname the#1\endcsname\quad}  
   {\csname #1@cntformat\endcsname}
}
\let\oldappendix\appendix 
\renewcommand\appendix{%
    \oldappendix
    \newcommand{\section@cntformat}{\appendixname~\thesection\quad}
}
\makeatother

\usepackage[dvipdfmx]{graphicx}

\usepackage{cite}
\usepackage[normalem]{ulem}
\usepackage{comment}

\usepackage[usenames,dvipsnames]{xcolor} 

\newtheorem{theorem}{Theorem}
\usepackage{amssymb,amsfonts,amsmath}
\usepackage[top=25truemm,bottom=25truemm,left=20truemm,right=20truemm]{geometry}
\usepackage{bm}
\usepackage{cases}
\usepackage[mathlines]{lineno}

\usepackage{setspace}

\title{Strategies that enforce linear payoff relationships under observation errors in Repeated Prisoner's Dilemma game}
\bigskip
\author{Azumi Mamiya${}^{1}$ and Genki Ichinose${}^{1*}$
\ \\
\ \\
${}^{1}$
Department of Mathematical and Systems Engineering, Shizuoka University, \\3-5-1 Johoku, Naka-ku, Hamamatsu, 432-8561, Japan\\
$^*$ Corresponding author (ichinose.genki@shizuoka.ac.jp)}

\begin{document}

\maketitle

\section*{Abstract}
The theory of repeated games analyzes the long-term relationship of interacting players and mathematically reveals the condition of how cooperation is achieved, which is not achieved in a one-shot game. 
In the repeated prisoner's dilemma (RPD) game with no errors, zero-determinant (ZD) strategies allow a player to unilaterally set a linear relationship between the player's own payoff and the opponent's payoff regardless of the strategy that the opponent implements. In contrast, unconditional strategies such as ALLD and ALLC also unilaterally set a linear payoff relationship.
Errors often happen between players in the real world.
However, little is known about the existence of such strategies in the RPD game with errors. Here, we analytically search for all strategies that enforce a linear payoff relationship under observation errors in the RPD game. As a result, we found that, even in the case with observation errors, the only strategy sets that enforce a linear payoff relationship are either ZD strategies or unconditional strategies and that no other strategies can enforce it, which were numerically confirmed.

\section*{Keywords}
Prisoner's dilemma, repeated games, observation errors, zero-determinant strategies, unconditional strategies

\section{Introduction\label{sec:introduction}}
The two-player repeated prisoner's dilemma (RPD) game is a model for exploring the long-term relationships of players, which mathematically reveals how cooperation and competition arise among competitive players \cite{MailathSamuelson2006book}.
In the one-shot PD game, defection is the only Nash equilibrium.
On the other hand, cooperation is possible in the RPD game because players can reward cooperating partners by cooperating in the future.
Also, players can punish defecting partners by defecting in the future.
This mechanism is called direct reciprocity \cite{Trivers1971QRevBiol, Nowak2006book, Sigmund2010book} and makes it possible for players to mutually cooperate in the RPD game.
In the context of the RPD game, theoretical biologists are interested in which strategies win in evolving populations. This question falls into the field of evolutionary games \cite{MaynardSmith1982book}.
A series of the results of evolutionary games in the RPD game brought promising findings.
Especially, with noise, generous tit-for-tat \cite{Nowak1992Nature-gtft} and win-stay lose-shift \cite{Nowak1993Nature, Kraines1993TheorDecis} were robust to various kinds of evolutionary opponents. In this way, theoretical biologists have traditionally focused on strong strategies obtained from evolutionary consequences.
However, we can ask a question from a different point of view: Are there any strategies which always win against the opponent irrespective of the opponent's strategy? Answering this question fosters greater understanding of the RPD game.

In 2012, Press and Dyson suddenly answered this question by finding a novel class of strategies which contain such ultimate strategies, called zero-determinant (ZD) strategies \cite{Press2012PNAS}. ZD strategies impose a linear relationship between the payoffs for a focal player and his opponent regardless of the strategy that the opponent implements. The discovery of ZD strategies inspired 
various relevant studies, including  their evolution \cite{Stewart2012PNAS,Akin2016ErgodicTheory,Adami2013NatComm,Hilbe2013PNAS,Hilbe2013PlosOne-zd,ChenZinger2014JTheorBiol,Szolnoki2014PhysRevE-zd,Szolnoki2014SciRep-zd,WuRong2014PhysRevE-zd,Hilbe2015JTheorBiol,LiuLi2015PhysicaA-zd,Xu2017PhysRevE-zd, Wang2019Chaos, Stewart2013PNAS, Mao2018EPL, Xu2019Neurocomputing}, 
multiplayer games \cite{Hilbe2014PNAS-zd,Hilbe2015JTheorBiol,Pan2015SciRep-zd,Milinski2016NatComm,Stewart2016PNAS}, continuous action spaces \cite{Mcavoy2016PNAS,Milinski2016NatComm,Stewart2016PNAS,Mcavoy2017TheorPopulBiol}, alternating games \cite{Mcavoy2017TheorPopulBiol}, animal contests \cite{EngelFeigel2018PhysRevE}, human reactions to computerized ZD strategies \cite{Hilbe2014NatComm,Wang2016NatComm-zd},
and human-human experiments \cite{Hilbe2016PlosOne,Milinski2016NatComm, Becks2019NatComm}, which promote an understanding of the nature of human cooperation. For further understanding, see the recent elegant classification of strategies, partners (called ``good strategies'' in Ref.~\cite{Akin2016ErgodicTheory, Akin2015Games}) and rivals, in direct reciprocity \cite{Hilbe2018NatHumBehav}.
In contrast, unconditional strategies such as ALLC and ALLD can also unilaterally set a linear payoff relationship against the opponent \cite{Hilbe2013PlosOne-zd, IchinoseMasuda2018JTheorBiol}.
A previous study revealed that those two types of strategies are the only sets which enforce a linear payoff relationship in the RPD game \cite{IchinoseMasuda2018JTheorBiol}.

These two types of strategies were found in the case of no errors.
Errors (or noise) are unavoidable in human interactions and they may lead to the collapse of cooperation due to negative effects.
Thus, the effect of errors has been considered in the literature of the RPD game \cite{Kandori2002JEconTheor, Nowak1995JMathBiol, Fudenberg2012AmEconRev, Fudenberg1994AmEconRev, Sekiguchi1997JEconTheor, Barlo2009JEconTheor, Mailath2002JEconTheor, Mailath2011GEB, Hilbe2017PNAS}.
However, except for \cite{Hao2015PhysRevE}, the effect of errors has not been considered for strategies that enforce a linear payoff relationship.
There are typically two types of errors: perception errors \cite{Fudenberg2012AmEconRev} and implementation errors \cite{Fudenberg1994AmEconRev}.
Hao et al.~considered the former case of the errors where players may misunderstand their opponent's action because the players can only rely on their private monitoring \cite{Kandori2002JEconTheor, Sekiguchi1997JEconTheor} instead of their opponent's direct action.
They remarkably showed that ZD strategies can exist even in the case that such observation errors are incorporated \cite{Hao2015PhysRevE}.
In their model, they mathematically searched for one of the cases where determinants become zero in line with Press and Dyson's formalism \cite{Press2012PNAS}.
More specifically, they only searched for the case where the second and fourth columns of the determinant take the same value as Press and Dyson did in the case of no errors.
They did not consider other possible strategies that make the determinant zero in the case of errors.
In this study, from all possibilities, we mathematically searched for all of the cases where the determinant becomes zero.
As a result, we found that only ZD strategies \cite{Press2012PNAS} and unconditional strategies \cite{Hilbe2013PlosOne-zd, IchinoseMasuda2018JTheorBiol}
are the two types which enforce a linear payoff relationship and that no other strategies exist to make the determinant zero.
We also confirmed this result by numerical calculations.

\section{Model}
We consider the symmetric two-person RPD game with observation errors in line with the previous studies \cite{Sekiguchi1997JEconTheor, Hao2015PhysRevE}. Each player $i \in\{X,Y\}$ chooses an action $a_i \in \{\rm C,\rm D\}$. Each player cannot see what action the opponent chose. Instead, they can only observe a signal $\omega_i \in \{g,b\}$, where $g$ and $b$ denote good and bad signals, respectively. The signal cannot be observed by the other
player, meaning that the signal is private information. 
Each player's signal $\omega_i$ basically depends on the opponent's action but is also affected by noise from the environment, which is a stochastic variable.
In other words, a player observes $g$ (or $b$) when the other player chooses an action C (or D).
However, when an error occurs, a player observes $b$ (or $g$) although the other player chooses an action C (or D) due to observation errors.
We define $\sigma (\omega | a)$ as the probability that a signal profile $\omega = (\omega_X,\omega_Y)$ is realized, given that an action profile $a=(a_X,a_Y)$ occurs. 
Let $\epsilon$ be the probability that an error happens to one particular player but not to the other and $\xi$ be the probability that an error happens to both players.
Then, the probability that an error occurs to neither player is $1-2\epsilon-\xi$.  For example, when both players choose action C, we have $\sigma(g,g|\rm C,\rm C)=1-2\epsilon-\xi$, $\sigma(b, g|\rm C,\rm C)=\sigma(\it{g, b}|\rm C,\rm C)=\epsilon$, and $\sigma(b, b|\rm C,\rm C)=\xi$. 
The realized payoff for each player depends only on the action he chose and the signal he received, which is denoted by $u_i(a_i,\omega_i)$. Let $u_i(\rm{C}, \it{g})$, $u_i(\rm{C}, \it{b})$, $u_i(\rm{D}, \it{g})$, and $u_i(\rm{D},\it{b})$ be $R,S,T$, and $P$, respectively. Then the payoff matrix is given by
\begin{equation}
\bordermatrix{
 &  g &  b \cr
{\rm C} & R & S \cr
{\rm D} & T & P \cr}.
\label{eq:payoff}
\end{equation}
The entries represent the payoffs that a focal player gains in a single round of the repeated game.
Each row and column represents the action that the focal player chose and the signal he observed, respectively.
In each stage, player $i$'s expected payoff value over all possible signals, when two players have an action profile $a$, is represented by
\begin{equation}
 f_i(a)=\sum_\omega u_i(a_i,\omega_i)\sigma(\omega|a).
 \label{cond_err}
\end{equation}
The expected payoffs under different action profiles $(\rm C,\rm C)$,$(\rm C,\rm D)$,$(\rm D,\rm C)$, and $(\rm D,\rm D)$ are denoted by $R_E$, $S_E$, $T_E$ and $P_E$, respectively. According to Eq.~\eqref{cond_err}, $R_E$, $S_E$, $T_E$, and $P_E$ are derived as $R_E=R (1-\epsilon-\xi)+S (\epsilon+\xi)$, $S_E=S(1-\epsilon-\xi)+R(\epsilon+\xi)$, $T_E=T(1-\epsilon-\xi)+P(\epsilon+\xi)$, $P_E=P(1-\epsilon-\xi)+T(\epsilon+\xi)$, respectively.
We assume that
\begin{equation}
T_E>R_E>P_E>S_E,
\label{eq:T>R>P>S}
\end{equation}
which dictates the prisoner's dilemma condition. Both players expect a larger payoff by selecting D rather than C irrespective of the other's action because $T_E>R_E$ and $P_E>S_E$ hold. We also assume that
\begin{equation}
2R_E>T_E+S_E,
\label{eq:2R>T+S}
\end{equation}
which guarantees that mutual cooperation is more beneficial than the two players alternating C and D in the opposite phase, i.e., CD, DC, CD, DC, $\ldots$, where the first and second letter represent the actions selected by $X$ and $Y$, respectively.
The two players repeat the game whose payoff matrix in each round is given by
Eq.~\eqref{eq:payoff}. 

Consider two players $X$ and $Y$ that adopt memory-one strategies, with which they use only the outcomes of the last round to decide the action to be submitted in the current round. Even in the case of memory-$n$ strategies, errors can be considered. In fact, Hilbe et al.~incorporated implementation errors in such a situation \cite{Hilbe2017PNAS}.
A memory-one strategy is specified by a 4-tuple; $X$'s strategy is given by a combination of
\begin{equation}
\bm{p}=(p_{\rm 1}, p_{\rm 2}, p_{\rm 3}, p_{\rm 4}),
\label{eq:def bm p}
\end{equation}
where $0\le p_{\rm 1}, p_{\rm 2}, p_{\rm 3}, p_{\rm 4}\le 1$. The subscripts 1, 2, 3, and 4 of $p$ mean previous outcome C$g$, C$b$, D$g$ and D$b$, respectively. In Eq.~\eqref{eq:def bm p}, $p_{\rm 1}$ is the conditional probability that $X$ cooperates when $X$ cooperated and observed signal $g$ in the last round, $p_{\rm 2}$ is the conditional probability that $X$ cooperates when $X$ cooperated and observed signal $b$ in the last round, $p_{\rm 3}$ is the conditional probability that $X$ cooperates when $X$ defected and observed signal $g$ in the last round, and $p_{\rm 4}$ is the conditional probability that $X$ cooperates when $X$ defected and observed signal $b$ in the last round. Note that, in this model, $\bm p$ depends on $X$'s action and its private observation in the last round \cite{Sekiguchi1997JEconTheor, Hao2015PhysRevE}. Contrary, $\bm p$ depends on $X$'s and $Y$'s direct actions in the last round in the case of no errors.
Similarly, $Y$'s strategy is specified by a combination of
\begin{equation}
\bm{q}=(q_{\rm 1}, q_{\rm 2}, q_{\rm 3}, q_{\rm 4}),
\end{equation}
where $0\le q_{\rm 1}, q_{\rm 2}, q_{\rm 3}, q_{\rm 4}\le 1$.
Because both players adopt a memory-one strategy, the stochastic state of the two players in round $t$ is described by $\bm v(t) = (v_1(t), v_2(t), v_3(t), v_4(t))$, where the subscripts 1, 2, 3, and 4 of $v$ mean the stochastic state (C,C), (C,D), (D,C), and (D,D), respectively.
$v_1(t)$ is the probability that both players cooperate in round $t$, $v_2(t)$ is the probability that $X$ cooperates and $Y$ defects in round $t$, and so forth.
The state transition matrix $M$ of this noisy repeated game is given by
\begin{equation}
\label{eq:M}
M=
\scalebox{0.85}{$\displaystyle
 \left(
 \begin{array}{ll}
    \left(
    \begin{array}{ll}
      \tau p_1 q_1  \\
      +\epsilon p_1 q_2 \\
      +\epsilon p_2 q_1 \\
      +\xi p_2 q_2 
    \end{array}
    \right)
    \left(
    \begin{array}{ll}
      \tau p_1 (1-q_1)  \\
      +\epsilon p_1 (1-q_2) \\
      +\epsilon p_2 (1-q_1) \\
      +\xi p_2 (1-q_2) 
    \end{array}
    \right)
    \left(
    \begin{array}{ll}
      \tau (1-p_1) q_1  \\
      +\epsilon (1-p_1) q_2 \\
      +\epsilon (1-p_2) q_1 \\
      +\xi (1-p_2) q_2 
    \end{array}
    \right)
    \left(
    \begin{array}{ll}
      \tau (1-p_1) (1-q_1)  \\
      +\epsilon (1-p_1) (1-q_2) \\
      +\epsilon (1-p_2) (1-q_1) \\
      +\xi (1-p_2) (1-q_2) 
    \end{array}
    \right)\\
  \left(
    \begin{array}{ll}
      \epsilon p_1 q_3  \\
      +\xi p_1 q_4 \\
      +\tau p_2 q_3 \\
      +\epsilon p_2 q_4 
    \end{array}
    \right)
    \left(
    \begin{array}{ll}
      \epsilon p_1 (1-q_3)  \\
      +\xi p_1 (1-q_4) \\
      +\tau p_2 (1-q_3) \\
      +\epsilon p_2 (1-q_4) 
    \end{array}
    \right)
    \left(
    \begin{array}{ll}
      \epsilon (1-p_1) q_3  \\
      +\xi (1-p_1) q_4 \\
      +\tau (1-p_2) q_3 \\
      +\epsilon (1-p_2) q_4 
    \end{array}
    \right)
    \left(
    \begin{array}{ll}
      \epsilon (1-p_1) (1-q_3)  \\
      +\xi (1-p_1) (1-q_4) \\
      +\tau (1-p_2) (1-q_3) \\
      +\epsilon (1-p_2) (1-q_4) 
    \end{array}
    \right)\\
    \left(
    \begin{array}{ll}
      \epsilon p_3 q_1  \\
      +\tau p_3 q_2 \\
      +\xi p_4 q_1 \\
      +\epsilon p_4 q_2 
    \end{array}
    \right)
    \left(
    \begin{array}{ll}
      \epsilon p_3 (1-q_1)  \\
      +\tau p_3 (1-q_2) \\
      +\xi p_4 (1-q_1) \\
      +\epsilon p_4 (1-q_2) 
    \end{array}
    \right)
    \left(
    \begin{array}{ll}
      \epsilon (1-p_3) q_1  \\
      +\tau (1-p_3) q_2 \\
      +\xi (1-p_4) q_1 \\
      +\epsilon (1-p_4) q_2 
    \end{array}
    \right)
    \left(
    \begin{array}{ll}
      \epsilon (1-p_3) (1-q_1)  \\
      +\tau (1-p_3) (1-q_2) \\
      +\xi (1-p_4) (1-q_1) \\
      +\epsilon (1-p_4) (1-q_2) 
    \end{array}
    \right)\\
    \left(
    \begin{array}{ll}
      \xi p_3 q_3  \\
      +\epsilon p_3 q_4 \\
      +\epsilon p_4 q_3 \\
      +\tau p_4 q_4 
    \end{array}
    \right)
    \left(
    \begin{array}{ll}
      \xi p_3 (1-q_3)  \\
      +\epsilon p_3 (1-q_4) \\
      +\epsilon p_4 (1-q_3) \\
      +\tau p_4 (1-q_4) 
    \end{array}
    \right)
    \left(
    \begin{array}{ll}
      \xi (1-p_3) q_3  \\
      +\epsilon (1-p_3) q_4 \\
      +\epsilon (1-p_4) q_3 \\
      +\tau (1-p_4) q_4 
    \end{array}
    \right)
    \left(
    \begin{array}{ll}
      \xi (1-p_3) (1-q_3)  \\
      +\epsilon (1-p_3) (1-q_4) \\
      +\epsilon (1-p_4) (1-q_3) \\
      +\tau (1-p_4) (1-q_4) 
    \end{array}
    \right)
 \end{array}
  \right)
$},
\end{equation}
where $\tau =1-2\epsilon-\xi$. Each row and column represents the previous states and the following states of the game, respectively.
Then, the stochastic state of the two players in round $t + 1$ is calculated by $\bm v(t + 1) = \bm v(t)M$.
The stationary distribution for $M$ is a vector $\bm v$ such that
\begin{equation}
\label{eq:vM}
\bm v=\bm v M.
\end{equation}
Eq.~\eqref{eq:vM} and $M^\prime \equiv M-I$ yield 
\begin{equation}
\label{eq:vMprime}
\bm v M^\prime=0.
\end{equation}
Applying Cramer's rule to matrix $M'$, we obtain
\begin{equation}
\label{eq:adjM}
Adj(M^\prime) M^\prime=0,
\end{equation}
where $Adj(M^\prime)$ is the adjugate matrix of $M^\prime$. Here, Eqs.~\eqref{eq:vMprime} and \eqref{eq:adjM} imply that every row of $Adj(M^\prime)$ is proportional to $\bm v$. Therefore, $\bm v$ is solely represented by the components of matrix $M^\prime$. Choosing the fourth row of the matrix $Adj(M^\prime)$, we see that $\bm v$ is composed of the determinant of the $3\times3$ matrixes formed from the first three columns of $M^\prime$.
We add the first column of $M^\prime$ into the second and third columns.
Even by this manipulation, this determinant is unchanged. 
The result of these manipulations is a formula for the dot product of an arbitrary vector $\bm f=(f_1,f_2,f_3,f_4)$ with the stationary distribution vector $\bm v$, which can be represented by the form of the determinant
\begin{equation}
\label{eq:D_err}
\begin{split}
  {\bm v} \cdot{\bm f}=
  & \left|
    \begin{array}{cccc}
      \tau p_1 q_1 +\epsilon p_1 q_2 +\epsilon p_2 q_1+\xi p_2 q_2 -1 & \mu p_{1}+\eta p_{2}-1 & \mu q_{1}+\eta q_{2}-1 & f_1\\
      \epsilon p_1 q_3+\xi p_1 q_4+\tau p_2 q_3+\epsilon p_2 q_4 & \eta p_{1}+\mu p_{2}-1 & \mu q_{3}+\eta q_{4} & f_2\\
      \epsilon p_3 q_1+\tau p_3 q_2+\xi p_4 q_1+\epsilon p_4 q_2 & \mu p_{3}+\eta p_{4} &  \eta q_{1}+\mu q_{2}-1 & f_3\\      
      \xi p_3 q_3+\epsilon p_3 q_4+\epsilon p_4 q_3+\tau p_4 q_4 & \eta p_{3}+\mu p_{4} & \eta q_{3}+\mu q_{4} & f_4
    \end{array}
  \right|
  \equiv D(\bm p,\bm q,\bm f),
\end{split}
\end{equation}
where $\mu=1-\epsilon-\xi$ and $\eta=\epsilon+\xi$.
If we replace the arbitrary vector $\bm f$ with $X$'s expected payoff vector $\bm S_X=(R_E,S_E,T_E,P_E)$, we obtain $\bm v \cdot \bm S_X$.
Then, we divide it by $\bm v \cdot \bm 1$.
Finally, we can obtain player $X$'s per-round expected payoff in the form of the determinant as follows: 
\begin{equation}\label{eq:sx2}
s_X=\frac{{\bm v} \cdot {\bm S_X}}{\bm v \cdot \bm1} 
   =\frac{D({\bm p,\bm q,\bm S_X})}{D({\bm p,\bm q,\bm 1})},
\end{equation}
where $\bm 1 =(1,1,1,1)$ is needed for the normalization. 
Similarly, player $Y$'s per-round payoff can be represented by the form of the determinant
\begin{equation}\label{eq:sy2}
s_Y=\frac{{\bm v} \cdot {\bm S_Y}}{\bm v \cdot \bm1} 
   =\frac{D({\bm p,\bm q,\bm S_Y})}{D({\bm p,\bm q,\bm 1})},
\end{equation}
where $\bm S_Y$ is $Y$'s expected payoff vector $(R_E,T_E,S_E,P_E)$.
Hereafter, we only consider the relationship between those two expected payoffs because they converge to certain expected values, respectively, if the stationary distributions exist in infinitely repeated games.
In contrast, other types of the payoff are worth investigating in finitely repeated games.

Moreover, we can consider the linear combination of $s_X$ and $s_Y$, which can be given by the form of the determinant
\begin{equation}
\label{D_linear_err}
\alpha s_X+\beta s_Y+\gamma 
=\frac{ D({\bm p,\bm q,\alpha \bm{S_X} +\beta \bm{S_Y}+\gamma \bm{1}})}{D({\bm p,\bm q,\bm 1})},
\end{equation}
where $\alpha, \beta$, and $\gamma$, are arbitrary constant. The numerator of the right side of Eq.~\eqref{D_linear_err} is expressed in the following:
\begin{equation}
\label{eq:D2_2}
\begin{split}
  & D(\bm p,\bm q,\alpha \bm S_X+\beta \bm S_Y+\gamma \bm 1)= \\
  & \left|
    \begin{array}{cccc}
      \tau p_1 q_1 +\epsilon p_1 q_2 +\epsilon p_2 q_1+\xi p_2 q_2 -1 &
      \mu p_{1}+\eta p_{2}-1 &
      \mu q_{1}+\eta q_{2}-1 &
      \alpha R_E +\beta R_E +\gamma \\
      \epsilon p_1 q_3+\xi p_1 q_4+\tau p_2 q_3+\epsilon p_2 q_4 &
      \eta p_{1}+\mu p_{2}-1 &
      \mu q_{3}+\eta q_{4} &
      \alpha S_E +\beta T_E +\gamma \\
      \epsilon p_3 q_1+\tau p_3 q_2+\xi p_4 q_1+\epsilon p_4 q_2 &
      \mu  p_3 +\eta p_4     &
      \eta q_1 +\mu  q_2 - 1 &
      \alpha T_E +\beta S_E +\gamma \\      
      \xi  p_3 q_3+\epsilon p_3 q_4+\epsilon p_4 q_3+\tau p_4 q_4 &
      \eta p_3 +\mu p_4 &
      \eta q_3 +\mu q_4 &
      \alpha P_E +\beta P_E +\gamma
    \end{array}
  \right|.
\end{split}
\end{equation}
If Eq.~\eqref{eq:D2_2} is zero, the relationship between the two players' payoffs becomes linear.
In the next section, we search for all of the solutions which satisfy this condition.

\section{Result}
We search for strategies that impose a linear relationship between the two players' payoffs regardless of their opponent's strategies in the RPD game with observation errors, which satisfy the following equation:
\begin{equation}
\alpha s_X + \beta s_Y + \gamma = 0.
\label{eq:pi_X and pi_Y linear}
\end{equation}
If the numerator of the right side of Eq.~\eqref{D_linear_err} is zero, Eq.~\eqref{eq:pi_X and pi_Y linear} holds.
In other words, if $D(\bm p,\bm q,\alpha \bm S_X+\beta \bm S_Y+\gamma \bm 1)=0$ is satisfied, there is a linear payoff relationship between the two players' payoffs.

Press and Dyson (without error) \cite{Press2012PNAS} and Hao et al.~(with error) \cite{Hao2015PhysRevE} only searched for the case that the second and fourth columns take the same value.
This makes the determinant become zero.
Here, from all possibilities, we search for all of the cases (including this case) that $D(\bm p,\bm q,\alpha \bm S_X+\beta \bm S_Y+\gamma \bm 1)=0$ holds.
The following determinant theorem gives such a condition.
\begin{theorem}\label{eq:douchi1}
\rm{For an $n\times n$ matrix A, the following holds:}
\begin{eqnarray*}
det(\rm A)=0 &\Leftrightarrow & \rm{The\ columns\ of\ matrix\ A\ are\ linearly \ dependent\ vectors}.
\end{eqnarray*}
\end{theorem}

We define $\bm a_i$ ($i \in \{1,2,3,4\}$) as $i$-th column vector of the determinant of Eq.~\eqref{eq:D2_2}. From the above theorem, if the columns of the determinant of Eq.~\eqref{eq:D2_2} are linearly dependent vectors, there exist real numbers $s,t,u,v,\alpha,\beta$, and $\gamma$, except for the trivial solution ($(s,t,u,v)=(0,0,0,0)$,$(\alpha,\beta,\gamma)=(0,0,0)$), such that
\begin{equation}\label{eq:subordination}
  s {\bm a_1}+t {\bm a_2}+ u {\bm a_3} +v {\bm a_4}=\bm 0,
\end{equation}
where vector $\bm 0$ denotes a zero vector. 

\subsection{Without errors (perfect monitoring)}
\subsubsection{Mathematical analysis}
In this section, we search for all of the strategies that enforce a linear payoff relationship without errors ($\epsilon=0$ and $\xi=0$). When there are no errors, the expected payoffs correspond to the original payoffs, i.e., $\bm S_X=(R_E,S_E,T_E,P_E)=(R, S,T,P)$ and $\bm S_Y=(R_E,T_E,S_E,P_E)=(R,T,S,P)$, respectively.
In addition, by substituting  $\epsilon=0$ and $\xi=0$ into Eq.~\eqref{eq:D2_2}, we obtain
\begin{equation}
\label{eq:D}
  D({\bm p,\bm q,\alpha \bm S_X+\beta \bm S_Y+\gamma \bm 1}) = 
  \left|
    \begin{array}{cccc}
      p_1 q_1 - 1  & p_1- 1 & q_1 - 1 & \alpha R+\beta R +\gamma \\
      p_2 q_3      & p_2- 1 & q_3     & \alpha S+\beta T +\gamma \\
      p_3 q_2      & p_3    & q_2 - 1 & \alpha T+\beta S +\gamma \\      
      p_4 q_4      & p_4    & q_4     & \alpha P+\beta P +\gamma 
    \end{array}
  \right|,
\end{equation}
which is the same with Press and Dyson's determinant \cite{Press2012PNAS}.
By the extensive calculations provided in Appendix \ref{appendix_no_error}, we found the only strategies that impose a linear payoff relationship between the two players' payoffs are either

\begin{equation}\label{eq:zd}
  \begin{split}
    p_1 - 1 &= \alpha R+\beta R+\gamma \\
    p_2 - 1 &= \alpha S+\beta T+\gamma \\
    p_3     &= \alpha T+\beta S+\gamma \\
    p_4     &= \alpha P+\beta P+\gamma,
  \end{split}
\end{equation}
or
\begin{equation}\label{eq:rrrr}
p_1=p_2=p_3=p_4.
\end{equation}

Equation \eqref{eq:zd} corresponds to ZD strategies without error (\cite{Press2012PNAS}, Eq.~(1) of \cite{Hilbe2013PlosOne-zd}, Eq.~(1) of  \cite{Hilbe2013PNAS}, and Eq.~(3) of  \cite{Hilbe2018NatHumBehav}).
Equation \eqref{eq:rrrr} is called unconditional strategies \cite{Hilbe2013PlosOne-zd}.
Only these strategy sets $\bm p$ can impose a linear relationship and no other strategies can impose it.

To conclude, in the RPD game under perfect monitoring, we showed that either ZD strategies or unconditional strategies can impose a linear relationship between the two players' payoffs. This is consistent with the previous result in the case with a discount factor but no errors \cite{IchinoseMasuda2018JTheorBiol}.

\subsubsection{Numerical examples}
\begin{figure}[t]
  \centering
  \includegraphics[width=0.9\columnwidth]{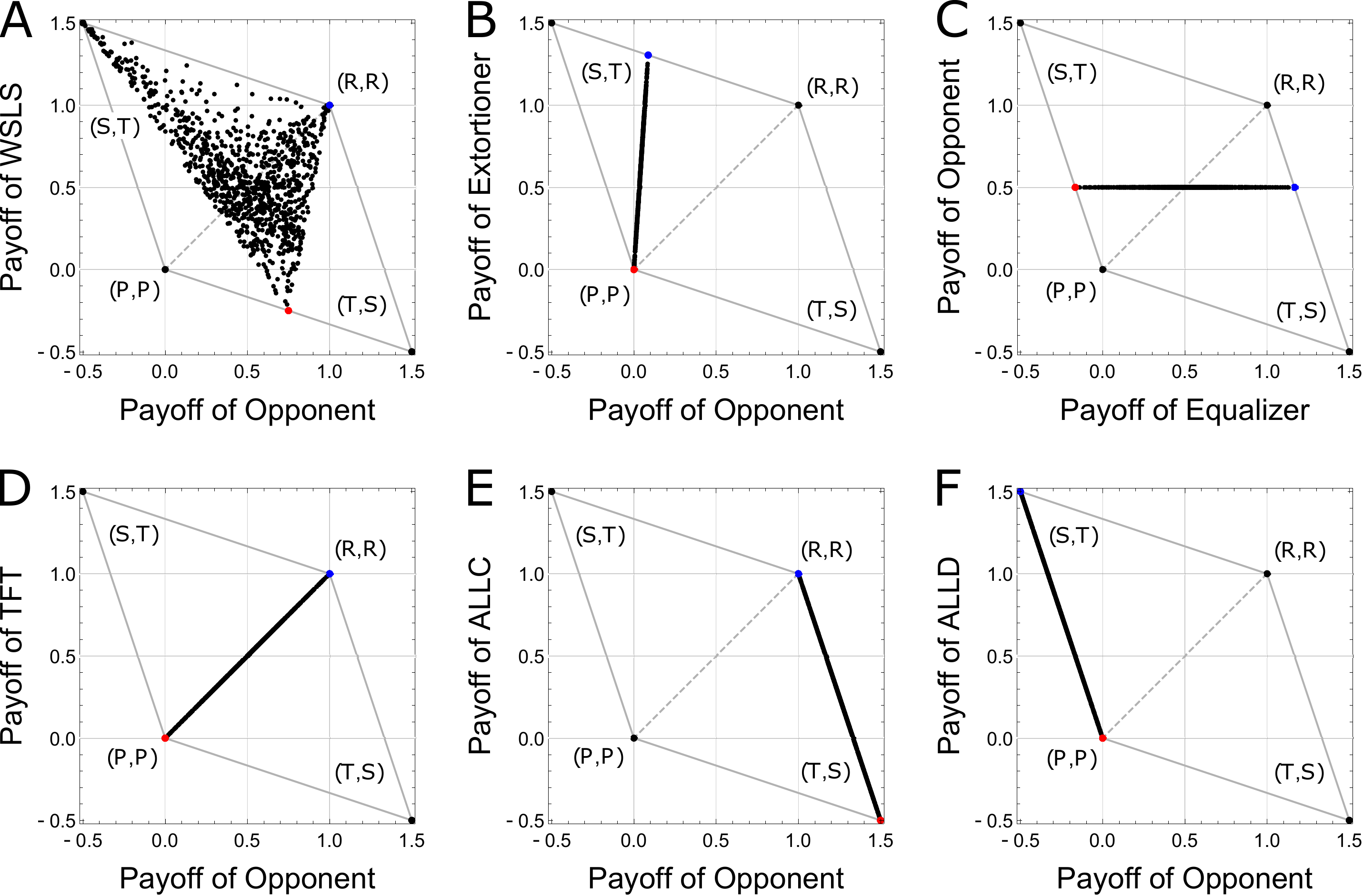}
  \caption{The payoff relationships between two players in the RPD game without errors. Payoff vector: $(T,R,P,S)=(1.5,1,0,-0.5)$.
  (A) WSLS strategy vs.~$1000+2$ strategies. (B) Extortioner strategy vs.~$1000+2$ strategies. (C) Equalizer strategy vs.~$1000+2$ strategies.
  (D) TFT strategy vs.~$1000+2$ strategies. (E) ALLC vs.~$1000+2$ strategies. (F) ALLD vs.~$1000+2$ strategies.}
  \label{figure1}
\end{figure}

We show numerical examples that ZD strategies and unconditional strategies can impose a linear relationship between the two players' payoffs while others cannot in the RPD game without errors.
Figure \ref{figure1} shows the relationship between the two players' expected payoffs per game with payoff vector $(T,R,P,S)=(1.5,1,0,-0.5)$.
The gray quadrangle in each panel represents the feasible set of the payoffs.
We fixed one particular strategy for player $X$ (vertical line) and randomly generate 1,000 strategies that satisfy $0 \leq q_1, q_2, q_3, q_4 \leq 1$ for player $Y$ (horizontal axis).
Thus, each black dot represents the payoff relationship between two players.
In addition, the blue and red are the particular cases for player $Y$. Red is the case that player $Y$ is ALLD and blue is the case that player $Y$ is  ALLC.

Figure \ref{figure1}A shows the case with a Win-Stay-Lose-Shift (WSLS) strategy vs.~$1000+2$ strategies.
As WSLS strategies are neither ZD nor unconditional strategies, the payoff relationships are not linear.

\subsubsection*{Numerical examples of ZD strategies}
Equalizer \cite{Press2012PNAS}, Extortioner \cite{Press2012PNAS}, and Generous strategies \cite{Stewart2013PNAS} are known as the three most prominent ZD strategies.
Here, we take up the first two as the numerical examples of ZD although Generous strategies play an important role in the evolution of cooperation.
In contrast to Extortion, Generous strategies always obtain lower payoffs than the opponent except for mutual cooperation.
Hence, Generous strategies are known as one of the cooperative ZD strategies.
Because Extortion never loses in a one-to-one competition, Extortion is feasible in a small population. However, in a large evolving population,
cooperative groups are more successful than the group of Extortioners. Thus, evolution leads from Extortion to Generous strategies \cite{Stewart2013PNAS}.
In this sense, Generous strategies are important.
Figure \ref{figure1}B is the case with an Extortioner strategy vs.~$1000+2$ strategies.
Extortioner strategies are the subset of ZD strategies \cite{Press2012PNAS} (See Box 1 in \cite{Hilbe2018NatHumBehav} for a clear explanation of Extortioner (extortionate) strategies).
Extortioner strategies can always gain a higher payoff than the one's opponent, except for the point $(P, P)$, regardless of the opponent's strategies.
When we set $(\alpha,\beta,\gamma)=(0.01,-0.15,0)$ in Eq.~\eqref{eq:zd}, we obtain an Extortioner strategy, $\bm p=(0.86, 0.77, 0.09, 0)$, with $0.01s_X-0.15s_Y=0$.
In this particular case, the Extortioner strategy (player $X$) gains the payoff fifteen times higher than player $Y$.

Figure \ref{figure1}C is the case with an Equalizer strategy vs.~$1000+2$ strategies. Note that, only in this case, the vertical and horizontal axes are reversed.
Thus, the horizontal axis is the payoff of Equalizer (player $X$) and the vertical axis is the payoff of player $Y$.
Equalizer strategies are also the subset of ZD strategies \cite{Press2012PNAS}.
If a player uses Equalizer strategies, he can fix the opponent's payoff to be one particular value.
When we set $(\alpha,\beta,\gamma)=(0,-2/3,1/3)$ in Eq.~\eqref{eq:zd}, we obtain an Equalizer strategy, $\bm p=(2/3,1/3,2/3,1/3)$, which can fix the opponent's payoff at $s_Y=0.5$ irrespective of the opponent's strategies.

Figure \ref{figure1}D is the case with TFT $\bm p=(1,0,1,0)$ strategy vs.~$1000+2$ strategies.
When we set $(\alpha,\beta,\gamma)=(0.5,-0.5,0)$  in Eq.~\eqref{eq:zd}, we obtain TFT $\bm p=(1,0,1,0)$, which means that TFT is also the subset of ZD strategies.
Actually, TFT is a special case of ZD strategies with $s_X=s_Y$ called ``fair strategies'' \cite{Hilbe2014PNAS-zd}.
Moreover, the strategies that $p_1=1, p_4=0, p_2+p_3=1$ including TFT can impose the linear payoff relationship $s_X=s_Y$.
See Appendix \ref{ap_2} for the proof.

\subsubsection*{Numerical examples of unconditional strategies}
Figure \ref{figure1}E is the case with ALLC vs.~$1000+2$ strategies.
ALLC is one of the examples of unconditional strategies $(r, r, r, r), 0 \leq r \leq 1$ where $r=1$.
If we substitute $r=1$ and $(T,R,P,S)=(1.5,1,0,-0.5)$ into Eq.~\eqref{rrrr_alpha_0}, we obtain $(\beta,\gamma)=(3\alpha,-4 \alpha)$ and we have a straight line represented by $s_X+3 s_Y-4=0$.
We numerically see that the payoff of ALLC is always lower than the opponent's payoff except for $(R, R)$.

Figure \ref{figure1}F is the case with ALLD vs.~$1000+2$ strategies.
ALLD is also one of the examples of unconditional strategies $(r, r, r, r), 0 \leq r \leq 1$ where $r=0$.
If we substitute $r=0$ and $(T,R,P,S)=(1.5,1,0,-0.5)$ into Eq.~\eqref{rrrr_alpha_0}, we obtain $(\beta,\gamma)=(3 \alpha,0)$ and we have a straight line represented by $s_X+3 s_Y=0$.
We numerically see that the payoff of ALLD is always higher than the opponent's payoff except for $(P, P)$.
Unlike ZD strategies, the slopes of the straight lines in Figure \ref{figure1}E and \ref{figure1}F are always negative \cite{Hilbe2013PlosOne-zd}.

\subsection{With observation errors (imperfect monitoring)}
\subsubsection{Mathematical analysis}
In the same way as no errors, we search for strategies that impose a linear relationship between the two players' payoffs regardless of the opponent's strategy in the RPD game with observation errors.
If the numerator of the right side of Eq.~\eqref{D_linear_err} is zero, the following equation holds:

\begin{equation}
\alpha s_X + \beta s_Y + \gamma = 0.
\label{eq:pi_X and pi_Y linear_error}
\end{equation}

In other words, if $D(\bm p,\bm q,\alpha \bm S_X+\beta \bm S_Y+\gamma \bm 1)=0$ is satisfied, there is a linear payoff relationship between the two players' payoffs.
By the extensive calculations provided in Appendix \ref{appendix_with_error}, we found the only strategies that impose a linear payoff relationship between the two players' payoffs are either

\begin{equation}\label{eq:hao}
\begin{split}
    \mu p_{1}+\eta p_{2}-1&= \alpha R_E+\beta R_E+\gamma \\
    \eta p_{1}+\mu p_{2}-1&= \alpha S_E+\beta T_E+\gamma \\
    \mu p_{3}+\eta p_{4}   &= \alpha T_E+\beta S_E+\gamma \\
    \eta p_{3}+\mu p_{4}   &= \alpha P_E+\beta P_E+\gamma,
\end{split}
\end{equation}
or
\begin{equation}\label{eq:rrrr2}
p_1=p_2=p_3=p_4.
\end{equation}

Equation \eqref{eq:hao} is ZD strategies with observation errors. This is consistent with Hao et al.'s \cite{Hao2015PhysRevE}.
Equation \eqref{eq:rrrr2} is unconditional strategies.
Moreover, we analytically show the feasible payoff range for unconditional strategies. See Appendix \ref{ap1}.

In summary, in the RPD game even with observation errors (imperfect monitoring), we showed that either ZD strategies or unconditional strategies can impose a linear relationship between the two players' payoffs and that no other strategies can impose it.
This is a new fact discovered in this study.

\subsubsection{Numerical examples}
\begin{figure}[t]
  \centering
  \includegraphics[width=0.9\columnwidth]{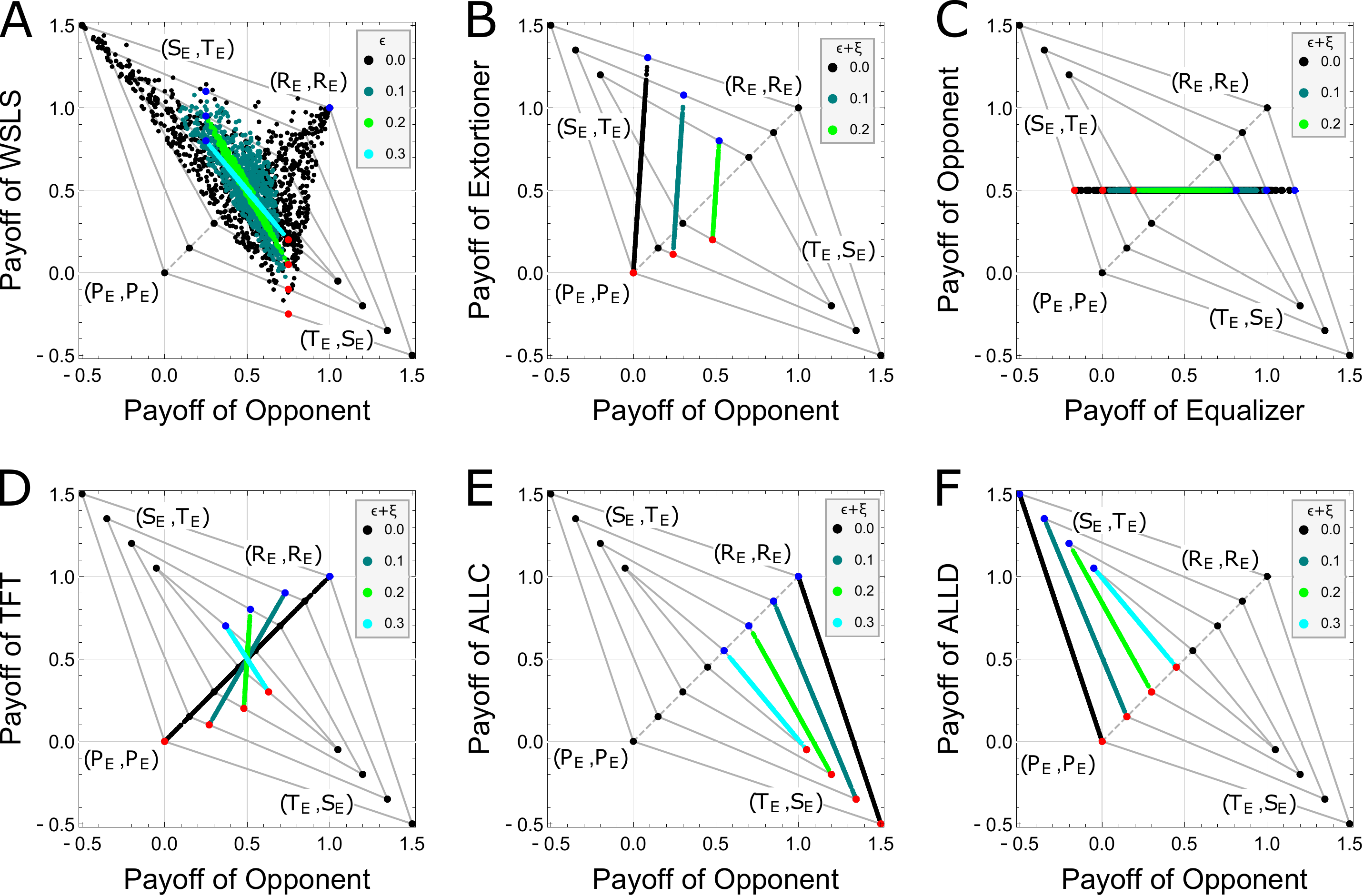}
  \caption{The payoff relationships between two players in the RPD game with errors. Payoff vector: $(T,R,P,S)=(1.5,1,0,-0.5)$.
  (A) WSLS strategy vs.~$1000+2$ strategies. (B) Extortioner strategy vs.~$1000+2$ strategies. (C) Equalizer strategy vs.~$1000+2$ strategies.
  (D) TFT strategy vs.~$1000+2$ strategies. (E) ALLC vs.~$1000+2$ strategies. (F) ALLD vs.~$1000+2$ strategies.}
  \label{figure2}
\end{figure}

As well as the case without errors, we show numerical examples that ZD strategies and unconditional strategies can impose a linear relationship between the two players' payoffs while others cannot in the RPD game with errors.
Figure \ref{figure2} shows the relationship between the two players' expected payoffs per game with payoff vector $(T,R,P,S)=(1.5,1,0,-0.5)$.
The gray quadrangle in each panel represents the feasible payoff set.
As error rates are increased, the size of the feasible payoff set becomes smaller.
We fixed one particular strategy for player $X$ (vertical line) and randomly generate 1,000 strategies that satisfy $0 \leq q_1, q_2, q_3, q_4 \leq 1$ for player $Y$ (horizontal axis).
Each black dot represents the payoff relationship between two players without errors ($\epsilon+\xi=0$), the same as Figure~\ref{figure1}.
Moreover, green, light green, and light blue dots correspond to the cases of  $\epsilon+\xi=0.1, 0.2$, and 0.3, respectively.
We do not consider the case of $\epsilon+\xi \ge 1/3$ because it does not satisfy the prisoner's dilemma condition: $T_E>R_E>P_E>S_E$. 
As in the case with no errors, red is the case that player $Y$ is ALLD and blue is the case that player $Y$ is  ALLC.

Figure~\ref{figure2}A shows the case with a Win-Stay-Lose-Shift (WSLS) strategy vs.~$1000+2$ strategies. In this case, $\xi=0$ is fixed and $\epsilon$ is varied to 0.1, 0.2 and 0.3.
As in the case with no errors, the payoff relationships are not linear in this case because WSLS strategies are neither ZD nor unconditional strategies.

\subsubsection*{Numerical examples of ZD strategies}
Figure~\ref{figure2}B is the case with an Extortioner strategy vs.~$1000+2$ strategies.
As shown in Figure~\ref{figure1}, $\bm p=(0.86, 0.77, 0.09, 0)$ (black dots) is the extortion strategy without errors.
In this case, player $X$ can always gain a higher payoff than the opponent (with the slope of 15), except for the point $(P, P)$, regardless of the opponent's strategies.
$\bm p=(0.926875, 0.818125, 0.111875, 0.003125)$ (green) and $\bm p=(1,0.86, 0.14, 0)$ (light green) are the extortion strategies when $\epsilon+\xi=0.1$ and 0.2, respectively.
Unlike Extortioner without errors, there exists the region that the expected payoff of the Extortioner with errors is lower than the opponent's payoff near $(P_E, P_E)$.

Hao et al. already proved this fact \cite{Hao2015PhysRevE}.
They call it \textit{dominant extortion} when the expected payoff of a focal player is always higher than the opponent except for  $(P, P)$.
This is only possible when there are no errors.
When there are errors, only \textit{contingent extortion} can exist as Hao et al. proved.
We assume that player $X$ adopts the contingent extortion.
The contingent extortion implies that when player $Y$ tries to increase his payoff, he will increase $X$'s payoff even more.
However, in some regions near $(P_E, P_E)$, $X$'s payoff is lower than $Y$'s payoff.
We mathematically restate the difference between dominant and contingent based on Hao et al.'s formalism \cite{Hao2015PhysRevE}.
We transform $\alpha=\phi s' ,\beta=-\phi,\gamma=\phi (1-s')l$ in Eq.~\eqref{eq:hao} in line with Hilbe's formalism \cite{Hilbe2013PlosOne-zd}.
We determine $l,s'$ so that $l=P_E+\Delta$，$1/s'>1$ are satisfied where $1/s'$ is the slope of the line.
Note that the inverse of $s'$ is considered as the slope because, in Hilbe's formalism, $s'$ is the coefficient for player $Y$ while in our and Hao's formalism $s'$ is the coefficient for player $X$.   
Also, $\phi,\Delta$ must satisfy $0 \leq p_1, p_2, p_3, p_4 \leq 1$.
When $\epsilon+\xi=0$ (no error), if we set $(s',\phi,\Delta)=(1/15,0.15,0)$, we obtain $l=P_E$ in Eq.~\eqref{eq:hao} and $\bm p$ becomes $\bm p=(0.86, 0.77, 0.09, 0)$ (black dots in Figure~\ref{figure2}B).
In this case, the payoff of player $X$ is always higher than player $Y$ except for the point $(P, P)$.
However, when $\epsilon+\xi > 0$, there is no solution in Eq.~\eqref{eq:hao} when $\Delta=0$.
Thus, $\Delta>0$ is needed, which means that there are the cases that the payoff of player $X$ is lower than that of player $Y$.
For instance, when $\epsilon+\xi=0.1, 0.2$ are given, if we set $(s',\phi,\Delta)=(1/15,0.15,0.1)$ and $(s',\phi,\Delta)=(1/15,0.15,0.2)$, $\bm p=(0.926875, 0.818125, 0.111875, 0.003125)$ (green in Figure~\ref{figure2}B) $\bm p=(1,0.86, 0.14, 0)$ (light green in Figure~\ref{figure2}B) are obtained.
In those cases, $X$'s payoff is lower than $Y$'s payoff near $(P_E, P_E)$ although $Y$'s increase leads to $X$'s increase even more.

Figure~\ref{figure2}C is the case with an Equalizer strategy vs.~$1000+2$ strategies. Note that, only in this case, the vertical and horizontal axes are reversed.
Thus, the horizontal axis is the payoff of Equalizer (player $X$) and the vertical axis is the payoff of player $Y$.
As Hao et al.~already suggested \cite{Hao2015PhysRevE}, there exist Equalizer strategies even if errors are incorporated.
When $\epsilon+\xi=0$ (no error), if we set $(\alpha,\beta,\gamma)=(0,-2/3,1/3)$ in Eq.~\eqref{eq:hao}, we obtain an Equalizer strategy, $\bm p=(2/3,1/3,2/3,1/3)$, which can fix the opponent payoff at $s_Y=0.5$ irrespective of the opponent's strategies as shown by black dots in Figure~\ref{figure2}C.
When $\epsilon+\xi=0.1$, if we set $(\alpha,\beta,\gamma)=(0,-0.695653,0.347827)$, we obtain an Equalizer strategy, $\bm p=(0.8,0.365217,0.634783,0.2)$ which can fix the opponent payoff at $s_Y=0.5$ as shown by green dots in Figure~\ref{figure2}C.
Also, when $\epsilon+\xi=0.2$, if we set $(\alpha,\beta,\gamma)=(0,-0.3,0.15)$, we obtain an Equalizer strategy, $\bm p=(0.99,0.74,0.26,0.01)$ which can fix the opponent payoff at $s_Y=0.5$ as shown by light green dots in Figure~\ref{figure2}C.
As error rates are increased, the payoff range for Equalizer becomes smaller.

Figure~\ref{figure2}D is the case with TFT $\bm p=(1,0,1,0)$ strategy vs.~$1000+2$ strategies.
When $\epsilon+\xi=0$ (no error), if we set $\bm p=(1,0,1,0)$ in Eq.~\eqref{eq:hao}, we obtain $(\alpha,\beta,\gamma)=(0.5,-0.5,0)$, which means that $s_X = s_Y$ (black dots) in the case of TFT.
When $\epsilon+\xi=0.1,0.2$, and 0.3, if we set $\bm p=(1,0,1,0)$ in Eq.~\eqref{eq:hao}, we obtain $(\alpha,\beta,\gamma)=(0.386555,-0.672269,0.142857)$, $(\alpha,\beta,\gamma)=(0.0714286,-1.07143,0.5)$, and $(\alpha,\beta,\gamma)=(-2.36364,-3.63636,3)$, respectively.
Thus, we obtain the corresponding lines, $(\alpha,\beta,\gamma)=(0.386555,-0.672269,0.142857)$ (green), $(\alpha,\beta,\gamma)=(0.0714286,-1.07143,0.5)$ (light green), and $(\alpha,\beta,\gamma)=(-2.36364,-3.63636,3)$ (light blue), respectively.
When there are no errors, $s_X = s_Y$ always holds. However, there are errors, this does not hold any more.
As error rates are increased, the difference between $s_X$ and $s_Y$ becomes larger.
In general, when there are errors, unlike when there are no errors, TFT does not enforce a linear payoff relationship. Only when special payoff matrices are given, the linear payoff relationship remains. 
See Appendix \ref{appendix_tft2} in detail.

\subsubsection*{Numerical examples of unconditional strategies}
Figure~\ref{figure2}E is the case with ALLC vs.~$1000+2$ strategies.
ALLC is one of the examples of unconditional strategies $(r, r, r, r), 0 \leq r \leq 1$ where $r=1$.
When $\epsilon+\xi=0$ (no error), by Eq.~\eqref{rrrr_alpha}, we obtain $(\beta,\gamma)= (3\alpha ,-4 \alpha)$.
Thus, the equation of the straight line is $s_X+3 s_Y -4=0$ (black dots in Figure~\ref{figure2}E) and the domain of $s_X$ becomes $-0.5 \leq s_X \leq 1$ from Eq.~\eqref{range}.
When $\epsilon+\xi=0.1,0.2$, and 0.3, we obtain the corresponding lines, $s_X+2.4 s_Y-2.89=0 \,(-0.35 \leq s_X \leq 0.85)$ (green)，$s_X+1.8 s_Y-1.96=0 \, (-0.2 \leq s_X \leq 0.7)$ (light green)，and $s_X+1.2 s_Y-1.21=0 \,(-0.05 \leq s_X \leq 0.55)$ (light blue), respectively.
We numerically see that the payoff of ALLC is always lower than the opponent's payoff except for $(R_E, R_E)$ and all the dots are on the feasible lines $(R_E, R_E) - (T_E, S_E)$, respectively.

Figure~\ref{figure2}F is the case with ALLD vs.~$1000+2$ strategies.
ALLD is also one of the examples of unconditional strategies $(r, r, r, r), 0 \leq r \leq 1$ where $r=0$.
When $\epsilon+\xi=0$ (no error), by Eq.~\eqref{rrrr_alpha}, we obtain $(\beta,\gamma)= (3\alpha ,0)$.
Thus, the equation of the straight line is $s_X+3 s_Y=0$ (black dots in Figure~\ref{figure2}F) and the domain of $s_X$ becomes $0 \leq s_X \leq 1.5$ from Eq.~\eqref{range}.
When $\epsilon+\xi=0.1,0.2$, and 0.3, we obtain the corresponding lines, $s_X+2.4 s_Y-0.51=0 \,(0.15 \leq s_X \leq 1.35)$ (green)，$s_X+1.8 s_Y-0.84=0 \, (0.3 \leq s_X \leq 1.2)$ (light green)，and $s_X+1.2 s_Y-0.99=0 \,(0.45 \leq s_X \leq 1.05)$ (light blue), respectively.
We numerically see that the payoff of ALLD is always higher than the opponent's payoff except for $(P_E, P_E)$ and all the dots are on the feasible lines $(S_E, T_E) - (P_E, P_E)$, respectively.

\section{Conclusions}
We analyzed strategies that enforce linear payoff relationships under observation errors in the RPD game.
Press and Dyson firstly developed a new mathematical formalism for the expected payoffs of two players and found that if the second and fourth columns of the specific determinant take the same value, the determinant becomes zero, which implies the two players' expected payoffs become linear \cite{Press2012PNAS}.
Hao et al.~used the same linear algebra technique and extended it to the case with observation errors \cite{Hao2015PhysRevE}.
Here, not just the case where the second and fourth columns of the determinant take the same value, we searched for all of the strategies which make the determinant zero under observation errors.
As a result, we found that the only strategy sets that enforce a linear payoff relationship are either ZD strategies or unconditional strategies, which was consistent with the case of the RPD game with a discount factor \cite{IchinoseMasuda2018JTheorBiol}.
We confirmed that the solutions are correct by showing some numerical calculations.

Press and Dyson first discovered strategies that make the determinant for the expected payoffs zero by finding that the second and fourth columns of the determinant take the same value \cite{Press2012PNAS}.
They call these strategies ``zero-determinant strategies'' (original ZD strategies) and all subsequent studies also call them ``zero-determinant strategies.''
By searching for all possibilities, we found that not only these original ZD strategies but also unconditional strategies make the determinant zero with a different form and that no other strategies exist to make the determinant zero.
In this sense, strictly speaking, both the original ZD strategies and unconditional strategies may be called ``zero-determinant strategies.''

The original ZD strategies and the unconditional strategies are the only sets which impose a linear payoff relationship irrespective of the opponent strategies, not only in the case with a discount factor \cite{IchinoseMasuda2018JTheorBiol} but also in the case with observation errors as shown here.
This result suggests that, in any case, those two sets are the only types of strategies that enforce a linear payoff relationship between two players. 
To investigate the inference, one possible direction of future research is analyzing the case of the RPD game with a discount factor under observation errors.

\appendix
\setcounter{figure}{0}
\renewcommand{\thefigure}{\Alph{section}.\arabic{figure}}

\section{Detailed calculations without errors\label{appendix_no_error}}
We substitute the column vectors of the determinant of Eq.~\eqref{eq:D} into Eq.~\eqref{eq:subordination} to obtain
\begin{equation}
\label{eq:f_0}
s
    \left(
    \begin{array}{c}
      p_1 q_1-1 \\
      p_2 q_3 \\
      p_3 q_2 \\
      p_4 q_4
    \end{array}
    \right)
  +t
  \left(
    \begin{array}{c}
      p_1-1 \\
      p_2-1 \\
      p_3 \\
      p_4
    \end{array}
    \right)
    +u
    \left(
    \begin{array}{c}
      q_1-1 \\
      q_3 \\
      q_2-1 \\
     q_4
    \end{array}
    \right)
    + v
   \left(
    \begin{array}{c}
     \alpha R +\beta R+\gamma\\
     \alpha S +\beta T+\gamma\\
     \alpha T +\beta S+\gamma\\
     \alpha P +\beta P+\gamma
    \end{array}
    \right)=\bm 0.
\end{equation}
By taking out $\bm q$ in Eq.~\eqref{eq:f_0}, we obtain 
\begin{equation}
\label{eq:vector}
    \left(
    \begin{array}{c}
     (s p_1+u) q_1 \\
     (s p_2+u) q_3 \\
     (s p_3+u) q_2 \\
     (s p_4+u) q_4 
    \end{array}
    \right)
    +t
    \left(
    \begin{array}{c}
      p_1-1 \\
      p_2-1 \\
      p_3 \\
      p_4
    \end{array}
    \right)
    +
    \left(
    \begin{array}{c}
      -u-s\\
      0\\
      -u\\
      0
    \end{array}
    \right)
    +v
   \left(
    \begin{array}{c}
     \alpha R +\beta R+\gamma\\
     \alpha S +\beta T+\gamma\\
     \alpha T +\beta S+\gamma\\
     \alpha P +\beta P+\gamma
    \end{array}
    \right)=\bm 0.
\end{equation}
Here, we search for strategies which satisfy $D(\bm p,\bm q,\alpha \bm S_X+\beta \bm S_Y+\gamma \bm 1)=0$ irrespective of  $Y$'s strategy $\bm q$, meaning that Eq.~\eqref{eq:vector} must hold true irrespective of $\bm q$. Therefore, the coefficients of each element $\bm q$ in Eq.~\eqref{eq:vector} must equal to zero, that is, the following conditions are necessary:
\begin{eqnarray}
\label{eq:seq1_0}
  \begin{cases}
    s p_1+u = 0 & \\
    s p_2+u = 0 & \\
    s p_3+u = 0 & \\
    s p_4+u = 0.
  \end{cases}
\end{eqnarray}
When Eq.~\eqref{eq:seq1_0} holds, the first terms of Eq.~\eqref{eq:vector} are eliminated and we obtain
\begin{equation}
\label{eq:48}
t
    \left(
    \begin{array}{c}
      p_1-1 \\
      p_2-1 \\
      p_3 \\
      p_4
    \end{array}
    \right)
    +
    \left(
    \begin{array}{c}
      -u-s\\
      0\\
      -u\\
      0
    \end{array}
    \right)
    +v
   \left(
    \begin{array}{c}
     \alpha R +\beta R+\gamma\\
     \alpha S +\beta T+\gamma\\
     \alpha T +\beta S+\gamma\\
     \alpha P +\beta P+\gamma
    \end{array}
    \right)=\bm 0.
\end{equation}
If there exist real numbers, $s, t, u,v, \alpha, \beta$, and $\gamma$ such that Eq.~\eqref{eq:seq1_0} and Eq.~\eqref{eq:48} are satisfied simultaneously, $D(\bm p,\bm q,\alpha \bm S_X+\beta \bm S_Y+\gamma \bm 1)=0$ holds irrespective of $\bm q$.
To solve Eq.~\eqref{eq:seq1_0}, we subtract the fourth equation from the first three in Eq.~\eqref{eq:seq1_0}:
\begin{eqnarray}
\label{eq:seq1_01}
  \begin{cases}
    s (p_1- p_4) &= 0 \\
    s (p_2- p_4) &= 0 \\
    s (p_3- p_4) &= 0 \\
    s p_4+u &= 0.
  \end{cases}
\end{eqnarray}
Then, we obtain $s=0$ or $p_1=p_4$ from the first equation.
First, in the case that $s=0$ holds, the second and third equations automatically hold and we obtain $u=0$ from the fourth. Hence, we obtain $s=0$ and $u=0$. Second, in the cases that $s\neq0$ and $p_1=p_4$ hold, we obtain $p_2=p_4$ and $p_3=p_4$ and $p_4=-u/s$ from the second, third and fourth equations, respectively. Therefore, the solutions of Eq.~\eqref{eq:seq1_0} are either  (1) $s=0$ and $u=0$ or  (2)  $p_1=p_2=p_3=p_4=-u/s$.
Next, we check that these solutions can also satisfy Eq.~\eqref{eq:48} in the following.

\subsubsection*{Case (1) $s=0$ and $u=0$:}
In this case, we substitute $s=0$ and $u=0$ into Eq.~\eqref{eq:48} to obtain
\begin{equation}
    t \left(
    \begin{array}{c}
     p_1 - 1 \\
     p_2 - 1 \\
     p_3     \\
     p_4
    \end{array}
    \right)+v
    \left(
    \begin{array}{c}
     \alpha R+\beta R+\gamma \\
     \alpha S+\beta T+\gamma \\
     \alpha T+\beta S+\gamma \\
     \alpha P+\beta P+\gamma
    \end{array}
    \right)=\bm 0.
\end{equation}
Here, when we set $t=0$, either equation
\begin{equation}
v=0
\end{equation}
or
\begin{equation}\label{alpha_beta_gamma_1}
    \left(
    \begin{array}{c}
     \alpha R+\beta R+\gamma \\
     \alpha S+\beta T+\gamma \\
     \alpha T+\beta S+\gamma \\
     \alpha P+\beta P+\gamma
    \end{array}
    \right)=\bm 0
\end{equation}
must hold. When we set $v=0$, we obtain the trivial solution $(s,t,u,v)=(0,0,0,0)$. Also, we solve Eq.~\eqref{alpha_beta_gamma_1} and obtain the trivial solution $(\alpha,\beta,\gamma)=(0,0,0)$. Hence, we do not have to consider the case of $t=0$.
Therefore, in the following,  we only consider $t\neq0$.
Replacing constants $-\alpha v/t$, $-\beta v/t$, and $-\gamma v/t$ with $\alpha$, $\beta$, and $\gamma$, we obtain, 
\begin{equation}\label{eq:zd_A}
  \begin{split}
    p_1 - 1 &= \alpha R+\beta R+\gamma \\
    p_2 - 1 &= \alpha S+\beta T+\gamma \\
    p_3     &= \alpha T+\beta S+\gamma \\
    p_4     &= \alpha P+\beta P+\gamma.
  \end{split}
\end{equation}
If there exist $\alpha,\beta$, and $\gamma$ for $\bm p$ satisfying Eq.~\eqref{eq:zd_A}, there must be solutions that Eq.~\eqref{eq:subordination} hold.
This strategy set $\bm p$ can impose a linear relationship. Eq.~\eqref{eq:zd_A} corresponds to ZD strategies without error (\cite{Press2012PNAS}, Eq.~(1) of \cite{Hilbe2013PlosOne-zd}, Eq.~(1) of  \cite{Hilbe2013PNAS}, and Eq.~(3) of  \cite{Hilbe2018NatHumBehav}).

\subsubsection*{Case (2) $p_1=p_2=p_3=p_4=-u/s$:}
In this case, let $r \ (0 \le r \le 1)$ be $-u/s$ , we substitute $p_1=p_2=p_3=p_4=r$ and $u=-s r$ into Eq.~\eqref{eq:48} to obtain
\begin{equation}\label{eq:vector_eqn}
t
    \left(
    \begin{array}{c}
     r-1\\
     r-1  \\
     r \\
     r
    \end{array}
    \right)+
    s \left(
    \begin{array}{c}
     r-1  \\
     0\\
     r\\
     0
    \end{array}
    \right)
    +v
    \left(
    \begin{array}{c}
     \alpha R+\beta R+\gamma \\
     \alpha S+\beta T+\gamma \\
     \alpha T+\beta S+\gamma \\
     \alpha P+\beta P+\gamma
    \end{array}
    \right)=\bm 0.
\end{equation}
There exist real numbers $s, t,u,v, \alpha, \beta$, and $\gamma$ which satisfies Eq.~\eqref{eq:vector_eqn} as follows:
\begin{equation}\label{rrrr_alpha_0}
\begin{split}
	s&=\frac{v\alpha (S(-P  - R  + S) + T(P + R  - T))} {(1-r)(P -S)+ r (T - R)} \\
	t&=\frac{v\alpha (S(2 P- S + r (- P - R + S)) +T (- 2 P  + T + r (P + R - T)))} {(1-r)(P -S)+ r (T - R)}\\
	u&=-sr\\
	\beta &= \frac{\alpha ((1-r)(T-P)+ r (R - S ))} { (1 - r)(P -S)+ r (T- R)}\\
	\gamma&=\frac{\alpha (S - T) ((-1 + r)^2 P + r (1 -r )(T+S)+ r^2 R)} {(1- r)(P-S) + r (T -R)}\\
	\forall& v, \alpha.
\end{split}
\end{equation}
Because there exist real numbers $s, t, u, v, \alpha, \beta$, and $\gamma$ such that Eq.~\eqref{eq:seq1_0}  and Eq.~\eqref{eq:48} are satisfied,
$p_1=p_2=p_3=p_4=r\ (0\le r \le 1)$ enforces a linear payoff relationship.
This strategy set is called unconditional strategies \cite{Hilbe2013PlosOne-zd}.
By transforming $\alpha, \beta, \gamma$ into $\alpha = \phi s^\prime,\beta=-\phi,\gamma=\phi(1-s^\prime)l$ in Eq.~\eqref{rrrr_alpha_0}, we obtain the following equations, which are the same as Eq.~(16) of \cite{Hilbe2013PlosOne-zd}:
\begin{equation}
\begin{split}
	l &=  (1-r)^2 P+r(1-r)(T+S)+r^2 R\\
	s^\prime &= -\frac{(1-r)(P-S)+r(T-R)}{(1-r)(T-P)+r(R-S)}\\
	\phi  &=(1-r)(T-P)+r(R-S).
\end{split}
\end{equation}

\section{Strategies that enforce $s_X=s_Y$ without errors}
\label{ap_2}
We prove that strategies specified by $p_1=1,p_4=0$ and $p_2+p_3=1$ including TFT enforce a linear payoff relationship with $s_X=s_Y$ under no errors.
Equation \eqref{eq:zd} can be rewritten as follows by transforming $\alpha, \beta, \gamma$ into $\alpha=\phi s' ,\beta=-\phi,\gamma=\phi (1-s')l$ where $s'$ is the slope of the straight line:
\begin{equation}\label{eq:zd-2}
  \begin{split}
    p_1  &=1 -\phi(1-s')(R-l) \\
    p_2  &=1 -\phi[s'(l-S)+(T-l)] \\
    p_3     &= \phi[(l-S)+s'(T-l)] \\
    p_4     &= \phi(1-s')(l-P),
  \end{split}
\end{equation}
which corresponds to Eq.~(16) of \cite{Hilbe2013PlosOne-zd}.
When $s'=1$, we obtain
\begin{equation}\label{eq:zd-s=1-1}
  \begin{split}
    p_1 \quad \quad &=1 \\
    p_2 +p_3 &=1  \\
    p_4 \quad \quad   &= 0.
  \end{split}
\end{equation}
This gives $(\alpha,\beta,\gamma)=(\phi,-\phi,0)$, hence, we obtain $s_X=s_Y$.
Thus, strategies specified by $p_1=1,p_4=0$ and $p_2+p_3=1$ enforce a linear payoff relationship with $s_X=s_Y$.

\section{Detailed calculations with errors\label{appendix_with_error}}
We substitute the column vectors of the determinant of Eq.~\eqref{eq:D2_2} into Eq.~\eqref{eq:subordination} to obtain
\begin{equation}
\label{eq:ff}
\begin{split}
   s \left(
    \begin{array}{c}
     \tau p_1 q_1 +\epsilon p_1 q_2 +\epsilon p_2 q_1+\xi p_2 q_2 -1 \\
     \epsilon p_1 q_3+\xi p_1 q_4+\tau p_2 q_3+\epsilon p_2 q_4 \\
     \epsilon p_3 q_1+\tau p_3 q_2+\xi p_4 q_1+\epsilon p_4 q_2 \\
     \xi p_3 q_3+\epsilon p_3 q_4+\epsilon p_4 q_3+\tau p_4 q_4
    \end{array}
    \right)
    +t
    \left(
    \begin{array}{c}
     \mu p_{1}+\eta p_{2}-1 \\
     \eta p_{1}+\mu p_{2}-1  \\
     \mu p_{3}+\eta p_{4} \\
     \eta p_{3}+\mu p_{4}
    \end{array}
    \right)
    & \\ +
    u
    \left(
    \begin{array}{c}
     \mu q_{1}+\eta q_{2}-1 \\
     \mu q_{3}+\eta q_{4}  \\
     \eta q_{1}+\mu q_{2}-1 \\
     \eta q_{3}+\mu q_{4}
    \end{array}
    \right)
   +v
   \left(
    \begin{array}{c}
     \alpha R_E+\beta R_E +\gamma \\
     \alpha S_E+\beta T_E +\gamma  \\
     \alpha T_E+\beta S_E +\gamma \\
     \alpha P_E+\beta P_E +\gamma
    \end{array}
    \right)=\bm 0.&
\end{split}
\end{equation}
By taking out $\bm q$ in Eq.~\eqref{eq:ff}, we obtain 
\begin{equation}
\label{eq:16}
\begin{split}
    \left(
    \begin{array}{c}
     (s(\tau p_1 +\epsilon p_2 ) +u\mu )q_1+(s(\epsilon p_1 +\xi p_2) +u\eta) q_2 \\
     (s(\epsilon p_1 +\tau p_2 ) +u\mu )q_3+(s(\xi p_1 +\epsilon p_2) +u\eta )q_4 \\
     (s(\epsilon p_3+\xi p_4) +u\eta )q_1+(s(\tau p_3+\epsilon p_4)  +u\mu) q_2\\
     (s(\xi p_3+\epsilon p_4 )+u\eta )q_3+(s(\epsilon p_3+\tau p_4) +u\mu )q_4
    \end{array}
    \right)
    +t 
    \left(
    \begin{array}{c}
     \mu p_{1}+\eta p_{2}-1\\
     \eta p_{1}+\mu p_{2}-1  \\
     \mu p_{3}+\eta p_{4} \\
     \eta p_{3}+\mu p_{4}
    \end{array}
    \right) 
   &\\
   +
    \left(
    \begin{array}{c}
     -s -u  \\
     0\\
     -u\\
     0
    \end{array}
    \right)
    +v
   \left(
    \begin{array}{c}
     \alpha R_E+\beta R_E +\gamma \\
     \alpha S_E+\beta T_E +\gamma  \\
     \alpha T_E+\beta S_E +\gamma \\
     \alpha P_E+\beta P_E +\gamma
    \end{array}
    \right)=\bm 0.&
\end{split}
\end{equation}
Here, we search for strategies which satisfy $D(\bm p,\bm q,\alpha \bm S_X+\beta \bm S_Y+\gamma \bm 1)=0$ irrespective of  $Y$'s strategy $\bm q$, meaning that Eq.~\eqref{eq:16} must hold true irrespective of $\bm q$. Therefore, the coefficients of each element $\bm q$ in Eq.~\eqref{eq:16} must equal to zero, that is, the following conditions are necessary:
\begin{eqnarray}\label{eq:seq1}
  \begin{cases}
    s(\epsilon p_1 +\xi p_2) +u\eta&= 0  \\
    s(\epsilon p_3+\xi p_4) +u\eta&= 0  \\
    s(\tau p_1 +\epsilon p_2 ) +u\mu &= 0  \\
    s(\tau p_3+\epsilon p_4)  +u\mu&= 0  \\
    s(\epsilon p_1 +\tau p_2 ) +u\mu& = 0  \\
    s(\xi p_1 +\epsilon p_2) +u\eta &= 0  \\
    s(\xi p_3+\epsilon p_4 )+u\eta&= 0  \\
    s(\epsilon p_3+\tau p_4) +u\mu&= 0 
  \end{cases}
\end{eqnarray}
When Eq.~\eqref{eq:seq1} holds, the first terms of Eq.~\eqref{eq:16} are eliminated and we obtain
\begin{equation}
\label{eq:17}
\begin{split}
t 
    \left(
    \begin{array}{c}
     \mu p_{1}+\eta p_{2}-1\\
     \eta p_{1}+\mu p_{2}-1  \\
     \mu p_{3}+\eta p_{4} \\
     \eta p_{3}+\mu p_{4}
    \end{array}
    \right) 
   +
    \left(
    \begin{array}{c}
     -s -u  \\
     0\\
     -u\\
     0
    \end{array}
    \right)
    +v
   \left(
    \begin{array}{c}
     \alpha R_E+\beta R_E +\gamma \\
     \alpha S_E+\beta T_E +\gamma  \\
     \alpha T_E+\beta S_E +\gamma \\
     \alpha P_E+\beta P_E +\gamma
    \end{array}
    \right)=\bm 0&.
\end{split}
\end{equation}
If there exist real numbers, $s, t, u, v, \alpha, \beta$, and $\gamma$ such that Eq.~\eqref{eq:seq1} and Eq.~\eqref{eq:17} are satisfied simultaneously, $D(\bm p,\bm q,\alpha \bm S_X+\beta \bm S_Y+\gamma \bm 1)=0$ holds irrespective of $\bm q$.
To solve Eq.~\eqref{eq:seq1}, we subtract the sixth equation from the first, the seventh from the second, the fifth from the third, and the eighth from the fourth  in Eq.~\eqref{eq:seq1} to obtain:
\begin{eqnarray}
\label{eq:seq}
  \begin{cases}
  s(\epsilon -\xi)(p_1 - p_2) &= 0  \\
    s(\epsilon-\xi)(p_3 - p_4)&= 0\\
    s(1-3 \epsilon -\xi)(p_1-p_2) &= 0  \\
    s(1-3 \epsilon -\xi)(p_3-p_4)&= 0  \\
    s(\epsilon p_1 +\tau p_2) +u\mu &= 0  \\
    s(\xi p_1 +\epsilon p_2) +u\eta &= 0  \\
    s(\xi p_3+\epsilon p_4 )+u\eta&= 0  \\
    s(\epsilon p_3+\tau p_4) +u\mu&= 0.
  \end{cases}
\end{eqnarray}
First, we solve the first four equations and obtain (1) $s=0$, (2) $\epsilon-\xi=0$ and $1-3 \epsilon -\xi=0$, (3) $p_1 - p_2=0$ and $p_3-p_4=0$.
We further analyze whether these equations satisfy the last four equations and Eq.~\eqref{eq:17} by dividing into three cases as follows.

\subsubsection*{Case (1) $s=0$:}
In this case, we substitute $s=0$ into Eq.~\eqref{eq:seq} to obtain
\begin{eqnarray}
\label{eq:20}
  \begin{cases}
    u\eta&= 0  \\
    u\mu&= 0, 
  \end{cases}
\end{eqnarray}
where $\mu=1-\epsilon-\xi$ and $\eta =\epsilon+\xi$. The equations $\mu=0$ and $\eta=0$ do not hold at the same time.
Therefore one of the solutions of Eq.~\eqref{eq:seq} is $s=0$ and $u=0$. Next, we check whether this solution satisfies Eq.~\eqref{eq:17}.
We substitute $s=0$ and $u=0$ into Eq.~\eqref{eq:17} to obtain
\begin{equation}
\begin{split}
t 
    \left(
    \begin{array}{c}
     \mu p_{1}+\eta p_{2}-1\\
     \eta p_{1}+\mu p_{2}-1  \\
     \mu p_{3}+\eta p_{4} \\
     \eta p_{3}+\mu p_{4}
    \end{array}
    \right) 
    +v
   \left(
    \begin{array}{c}
     \alpha R_E+\beta R_E +\gamma \\
     \alpha S_E+\beta T_E +\gamma  \\
     \alpha T_E+\beta S_E +\gamma \\
     \alpha P_E+\beta P_E +\gamma
    \end{array}
    \right)=\bm 0.&
\end{split}
\end{equation}
Here, when we set $t=0$, either equation
\begin{equation}
v=0
\end{equation}
or
\begin{equation}\label{alpha_beta_gamma_2}
    \left(
    \begin{array}{c}
     \alpha R_E+\beta R_E+\gamma \\
     \alpha S_E+\beta T_E+\gamma \\
     \alpha T_E+\beta S_E+\gamma \\
     \alpha P_E+\beta P_E+\gamma
    \end{array}
    \right)=\bm 0
\end{equation}
must hold. When we set $v=0$, we obtain the trivial solution $(s,t,u,v)=(0,0,0,0)$. Also, we solve Eq.~\eqref{alpha_beta_gamma_2} and obtain the trivial solution $(\alpha,\beta,\gamma)=(0,0,0)$. Hence, we do not have to consider the case of $t=0$.
Therefore, in the following,  we only consider $t\neq0$.
Replacing constants $-\alpha v/t$, $-\beta v/t$, and $-\gamma v/t$ with $\alpha$, $\beta$, and $\gamma$, we obtain, 
\begin{equation}\label{eq:hao_A}
\begin{split}
    \mu p_{1}+\eta p_{2}-1&= \alpha R_E+\beta R_E+\gamma \\
    \eta p_{1}+\mu p_{2}-1&= \alpha S_E+\beta T_E+\gamma \\
    \mu p_{3}+\eta p_{4}   &= \alpha T_E+\beta S_E+\gamma \\
    \eta p_{3}+\mu p_{4}   &= \alpha P_E+\beta P_E+\gamma.
\end{split}
\end{equation}
If there exist $\alpha,\beta$, and $\gamma$ satisfying Eq.~\eqref{eq:hao_A}, there must be solutions that Eq.~\eqref{eq:subordination} hold.
This solution is ZD strategies with errors. This is consistent with Hao et al.'s \cite{Hao2015PhysRevE}.

\subsubsection*{Case (2) $\epsilon-\xi=0$ and $1-3 \epsilon -\xi=0$:}
In this case, the equations $\epsilon-\xi=0$ and $1-3 \epsilon -\xi=0$ lead to $\epsilon=1/4$ and $\xi=1/4$.
When $\epsilon=1/4$ and $\xi=1/4$, the expected payoffs $R_E=1/2(R+S), S_E=1/2(R+S), T_E=1/2(T+P)$, and $ P_E=1/2(T+P)$ hold, which do not satisfy the condition of the prisoner's dilemma game: $T_E>R_E>P_E>S_E$.
Hence, we can exclude this solution.

\subsubsection*{Case (3) $p_1 - p_2=0$ and $p_3-p_4=0$:}
In this case, we substitute $p_1 - p_2=0$ and $p_3-p_4=0$ into Eq.~\eqref{eq:seq} to obtain
\begin{eqnarray}
  \begin{cases}
    \mu(s p_1 +u)&= 0  \\
    \eta(s  p_1 +u) &= 0  \\
    \eta(s p_3+u)&= 0  \\
    \mu(s p_3 +u)&= 0.
  \end{cases}
\end{eqnarray}
The equations $\mu=0$ and $\eta=0$ do not hold at the same time.
The following equations must hold.
\begin{eqnarray}
  \begin{cases}
    s p_1 +u &= 0  \\
    s p_3 +u &= 0 .
  \end{cases}
\end{eqnarray}
Therefore, we obtain the solution $p_1=p_2=p_3=p_4=-u/s$, which is the other solution of Eq.~\eqref{eq:seq}.
Let $r\ (0 \le r \le 1)$ be $-u/s$. Next, we check whether this solution satisfies Eq.~\eqref{eq:17}.
We substitute $p_1=p_2=p_3=p_4=r$ and $u=-s r$ into Eq.~\eqref{eq:17} to obtain
\begin{equation}\label{eq:vector_eqn2}
    t
    \left(
    \begin{array}{c}
     r-1\\
     r-1  \\
     r \\
     r
    \end{array}
    \right)+s \left(
    \begin{array}{c}
     r-1  \\
     0\\
     r\\
     0
    \end{array}
    \right)
    +v
    \left(
    \begin{array}{c}
     \alpha R_E+\beta R_E+\gamma \\
     \alpha S_E+\beta T_E+\gamma \\
     \alpha T_E+\beta S_E+\gamma \\
     \alpha P_E+\beta P_E+\gamma
    \end{array}
    \right)=\bm 0.
\end{equation}
There exist real numbers $s, t,u,v ,\alpha, \beta$, and $\gamma$ which satisfies Eq.~\eqref{eq:vector_eqn2} as follows:
\begin{equation}\label{rrrr_alpha}
\begin{split}
	s&=\frac{v\alpha (S_E(-P_E  - R_E  + S_E) + T_E(P_E + R_E  - T_E))} {(1-r)(P_E -S_E)+ r (T_E - R_E)}\\
	t&=\frac{v\alpha (S_E(2 P_E- S_E + r (- P_E - R_E + S_E)) +T_E (- 2 P_E  + T_E + r (P_E + R_E - T_E)))} {(1-r)(P_E -S_E)+ r (T_E - R_E)}\\
	u&=-sr\\
	\beta &= \frac{\alpha ((1-r)(T_E-P_E)+ r (R_E - S_E ))} { (1 - r)(P_E -S_E)+ r (T_E- R_E)}\\
	\gamma&=\frac{\alpha (S_E - T_E) ((-1 + r)^2 P_E + r (1 -r )(T_E+S_E)+ r^2 R_E)} {(1- r)(P_E-S_E) + r (T_E -R_E)}\\
	\forall& v,  \alpha.
\end{split}
\end{equation}
This strategy set is unconditional strategies $\bm p = (r,r,r,r), 0 \leq r \leq 1$.
Therefore, the unconditional strategies enforce a linear payoff relationship in the RPD game with errors because there exist real numbers $s, t, u, v, \alpha, \beta$, and $\gamma$ such that Eq.~\eqref{eq:seq1}  and Eq.~\eqref{eq:17} are satisfied.

\section{The feasible payoff-range for unconditional strategies}\label{ap1}
In this section, we show the feasible expected payoff-range when a player takes unconditional strategies. We assume that player $X$ takes unconditional strategies, which is $\bm p=(r,r,r,r)$. By substituting unconditional strategies $p_1=p_2=p_3=p_4=r$ into Eq.~\eqref{eq:D_err}, we obtain
\begin{equation}
  D({\bm p,\bm q,\bm f}) = 
  \left|
    \begin{array}{cccc}
      \tau  r q_1 +\epsilon r q_2 +\epsilon r q_1+\xi r q_2 -1 & \mu r+\eta r -1 & \mu q_{1}+\eta q_{2}-1 & f_1\\
      \epsilon r q_3+\xi r q_4+\tau r q_3+\epsilon r q_4 & \eta r +\mu r -1 & \mu q_{3}+\eta q_{4} & f_2\\
      \epsilon r q_1+\tau r q_2+\xi r q_1+\epsilon r q_2 & \mu r +\eta r &  \eta q_{1}+\mu q_{2}-1 & f_3\\      
      \xi r q_3+\epsilon r q_4+\epsilon r q_3+\tau r q_4 & \eta r +\mu r & \eta q_{3}+\mu q_{4} & f_4
    \end{array}
  \right|.
\end{equation}
The equations $\tau=1-2\epsilon-\xi,\mu=1-\epsilon -\xi$ and $\eta=\epsilon+\xi,\mu+\eta=1$ lead to
\begin{equation}
  D({\bm p,\bm q,\bm f}) = 
  \left|
    \begin{array}{cccc}
      r(\mu q_1 +\eta q_2) -1 & r-1 & \mu q_{1}+\eta q_{2}-1 & f_1\\
      r(\mu q_3 +\eta q_4) & r-1 & \mu q_{3}+\eta q_{4} & f_2\\
      r(\eta q_1+\mu q_2) & r &  \eta q_{1}+\mu q_{2}-1 & f_3\\      
      r(\eta q_3+\mu q_4)& r & \eta q_{3}+\mu q_{4} & f_4
    \end{array}
  \right|.
\end{equation}
By subtracting $r$ times the third column from the first, we obtain
\begin{equation}
  D({\bm p,\bm q,\bm f}) = 
  \left|
    \begin{array}{cccc}
      r-1 & r-1 & \mu q_{1}+\eta q_{2}-1 & f_1\\
      0 & r-1 & \mu q_{3}+\eta q_{4} & f_2\\
      r & r  &  \eta q_{1}+\mu q_{2}-1 & f_3\\      
      0 & r  & \eta q_{3}+\mu q_{4} & f_4
    \end{array}
  \right|.
\end{equation}
By subtracting the third row from the first and the fourth from the second, we obtain
\begin{equation}
  D({\bm p,\bm q,\bm f}) = \left|
    \begin{array}{cccc}
      -1 & -1   & (\mu - \eta) (q_1 - q_2)  & f_1-f_3\\
      0  & -1  & (\mu - \eta) (q_3 - q_4)  & f_2-f_4\\
      r  & r   & \eta q_1 +\mu q_2-1  & f_3\\      
      0  & r   & \eta q_3 +\mu q_4    & f_4
    \end{array}
  \right|.
\end{equation}
By subtracting $r$ times the first row from the third and the fourth from $r$ times the second, we obtain
\begin{equation}
   D({\bm p,\bm q,\bm f}) = \left|
    \begin{array}{cccc}
      -1 & -1   & (\mu - \eta) (q_1 - q_2)  & f_1-f_3\\
      0  & -1  & (\mu - \eta) (q_3 - q_4)  & f_2-f_4\\
      0  & 0   & \eta q_1 +\mu q_2-1 +r(\mu - \eta) (q_1 -q_2) & f_3+r(f_1-f_3)\\      
      0  & 0   & \eta q_3 +\mu q_4   +r(\mu - \eta) (q_3 -q_4) & f_4+r(f_2-f_4)
    \end{array}
  \right|.
\end{equation}
The Laplace expansion along the first column yields:
\begin{equation}
  D(\bm p,\bm q,\bm f)=-
  \left|
    \begin{array}{ccc}
      -1  & (\mu - \eta) (q_3 -q_4)  & f_2-f_4\\
      0   & \eta q_1 +\mu q_2-1 +r(\mu - \eta) (q_1 -q_2) & f_3+r(f_1-f_3)\\      
      0   & \eta q_3 +\mu q_4   +r(\mu - \eta) (q_3 -q_4) & f_4+r(f_2-f_4)
    \end{array}
  \right|.
\end{equation}
Additionally, the Laplace expansion along the first column yields:
\begin{equation}
  D({\bm p,\bm q,\bm f}) = 
  \left|
    \begin{array}{ccc}
       \eta q_1 +\mu q_2-1 +r(\mu - \eta) (q_1 -q_2) & r (f_1-f_3) + f_3\\      
       \eta q_3 +\mu q_4   +r(\mu - \eta) (q_3 -q_4) & r (f_2- f_4)+ f_4\\
    \end{array}
  \right|.
\end{equation}
Therefore $X$'s expected payoff can be calculated by the form of the determinant as follows:
\begin{equation}\label{eq:45}
\begin{split}
s_X&=\frac{{\bm v} \cdot {\bm S_X}}{\bm v \cdot \bm 1} =\frac{ D({\bm p,\bm q,\bm S_X})} {D({\bm p,\bm q,\bm 1})} \\
&=\frac{ \left|
    \begin{array}{ccc}
        \eta q_1 +\mu q_2-1 +r(\mu - \eta) (q_1 -q_2) & r (R_E-T_E) + T_E\\      
       \eta q_3 +\mu q_4   +r(\mu - \eta) (q_3 -q_4)& r (S_E- P_E)+ P_E\\
    \end{array}
  \right|} 
  {\left|
    \begin{array}{ccc}
       \eta q_1 +\mu q_2-1 +r(\mu - \eta) (q_1 -q_2)  & 1\\      
       \eta q_3 +\mu q_4   +r(\mu - \eta) (q_3 -q_4)& 1\\
    \end{array}
  \right|}.
\end{split}
\end{equation}
Let $x$ be $\eta q_1 +\mu q_2-1 +r(\mu - \eta) (q_1 -q_2)$ and $y$ be $\eta q_3 +\mu q_4   +r(\mu - \eta) (q_3 -q_4)$ to obtain
\begin{eqnarray}\label{eq:47}
s_X=\frac{ \left|
    \begin{array}{cc}
      x & r (R_E-T_E) + T_E\\
      y & r (S_E- P_E)+ P_E
    \end{array}
    \right|} {\left|
    \begin{array}{cc}
      x  & 1\\
      y  & 1
    \end{array}
    \right|} 
    =\frac{x \{ r (S_E- P_E)+ P_E\} -y \{ r (R_E-T_E) + T_E\}}{x - y},
\end{eqnarray}
where $-1\le x\le 0$ and $0\le y\le 1$ because $0\le q_1,q_2,q_3,q_4\le1$. In the case of $x\neq0$, let $k$ be $y/x$, where $-\infty<k\le0$ ($\because -1\le x< 0$ and $0\le y\le 1$). Then, Eq.~\eqref{eq:47} leads to
\begin{eqnarray}
s_X &=& \frac{ \{ r (S_E- P_E)+ P_E\} -k \{r (R_E-T_E) + T_E\}}{1 - k}\nonumber\\
      &=&\frac{ \{r (S_E- P_E)+ P_E\} - \{r (R_E-T_E) + T_E\}+(1-k) \{r (R_E-T_E) + T_E\}}{1 - k}  \nonumber \\
     &=& r (R_E-T_E) + T_E +\bigl[ \{r (S_E- P_E)+ P_E\} -\{r (R_E-T_E) + T_E\} \bigl]\frac{1}{1 - k}  \nonumber \\
     &=& r (R_E-T_E) + T_E - \{r (R_E-S_E)+(1- r) (T_E-P_E)\} \frac{1}{1 - k} .
\end{eqnarray}
Here, by the conditions $0\le r \le1$ and $T_E>R_E>P_E>S_E$, the $s_X$ is maximum if the function $f(k)=1/(1-k)$ is minimum and the $s_X$ is minimum if the function $f(k)$ is maximum. Then, the maximum of $f(k)$ is $f(0)=1$ and the minimum do not exist but $\lim_{k \to -\infty} f(k) \approx 0$. Hence, the range of $s_X$ in the case of $x\neq0$ is $ r (S_E- P_E)+ P_E \leq s_X < r (R_E-T_E) + T_E$. Next, in the case of $x=0$, $s_X=r (R_E-T_E) + T_E$ holds. From the above, the feasible expected payoff-range for unconditional strategies is given by
\begin{equation}
\label{range}
  r (S_E- P_E)+ P_E \leq s_X \leq  r (R_E-T_E) + T_E.
\end{equation}

For instance, when $(T, R, P, S)=(1.5, 1, 0, -0.5)$ is given, the expected payoffs become $R_E = 1-1.5(\epsilon + \xi), S_E=-0.5+1.5(\epsilon+\xi), T_E=1.5(1-\epsilon-\xi)$, and $P_E=1.5(\epsilon+\xi)$, respectively.
If player $X$ is ALLD ($r=0$), his expected payoff becomes $1.5(\epsilon+\xi) \leq s_X \leq  1.5(1-\epsilon-\xi)$ by Eq.~\eqref{range}.
Thus, when $\epsilon+\xi=0, 0.1, 0.2$, and 0.3, the ranges become $0 \leq s_X \leq  1.5$, $0.15 \leq s_X \leq  1.35$, $0.3 \leq s_X \leq  1.2$, and $0.45 \leq s_X \leq  1.05$, respectively.

Moreover, we can even know the possible payoff range for $s_Y$ when player $X$ takes $\bm p =(r,r,r,r)$ by replacing $\bm S_X$ with $\bm S_Y$ in Eq.~\eqref{eq:45}.

\section{TFT can enforce a linear payoff relationship under errors only when special conditions are satisfied}\label{appendix_tft2}
In general, with errors, TFT can enforce a linear payoff relationship only when special conditions are satisfied.
We prove it in this section.

As we showed in the main text, the only strategies that enforce a linear payoff relationship are either ZD or unconditional strategies with observation errors.
Thus, TFT must be included in one of them.
It is obvious that TFT is not classified as an unconditional strategy because it is specified by $\bm p=(1,0,1,0)$.
Therefore, we check whether TFT can be classified as ZD strategies.
If there exist strategies that satisfy Eq.~\eqref{eq:hao}, TFT is one of the ZD strategies.

\begin{figure}[t]
  \centering
  \includegraphics[width=0.4\columnwidth]{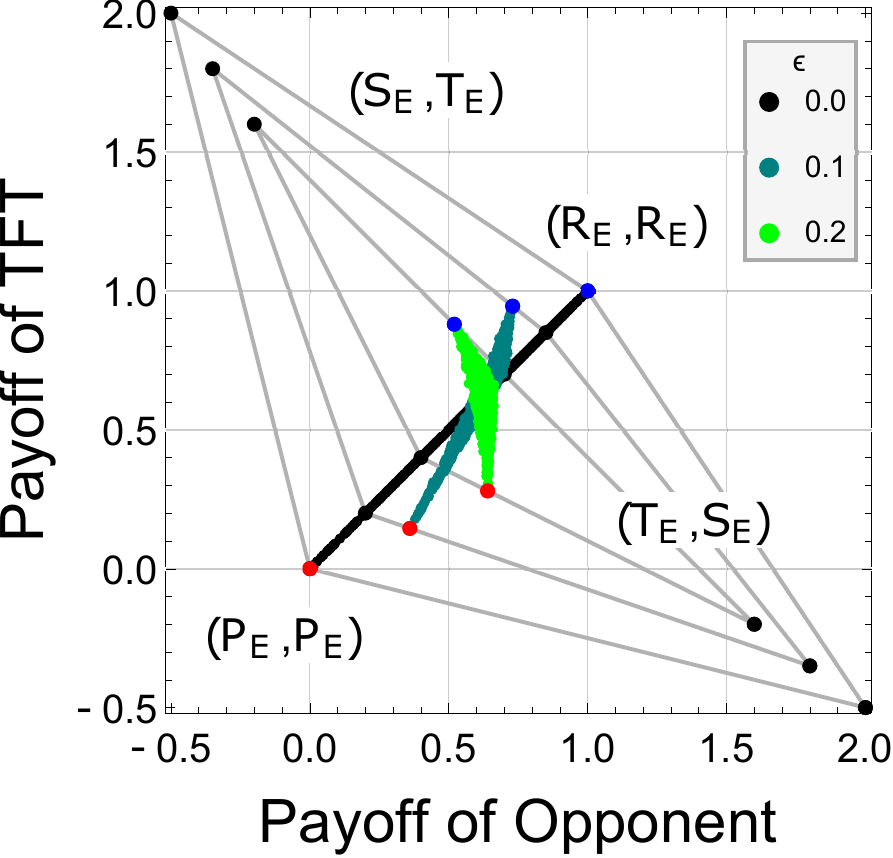}
  \caption{The payoff relationships between two players in the RPD game with errors. Payoff vector: $(T,R,P,S)=(2,1,0,-0.5)$.
  TFT strategy vs.~$1000+2$ strategies. In this case, $\xi=0$ is fixed and $\epsilon$ is varied to 0.1, 0.2 and 0.3.}
  \label{figure3}
\end{figure}

By substituting $\bm p =(1,0,1,0)$ into Eq.~\eqref{eq:hao}, we obtain
\begin{equation}\label{eq:hao_tft}
  \begin{split}
    \mu -1&= \alpha R_E+\beta R_E+\gamma \\
    \eta -1&= \alpha S_E+\beta T_E+\gamma \\
    \mu    &= \alpha T_E+\beta S_E+\gamma \\
    \eta    &= \alpha P_E+\beta P_E+\gamma. \\
  \end{split}
\end{equation}
We solve this equation and obtain the following two types of the solution:
\begin{equation}\label{Solution1}
  \begin{split}
   \eta&=0\\
    \alpha  &= \frac{-1}{S_E-T_E} \\
    \beta    &=  \frac{1}{S_E-T_E} \\
    \gamma   &= 0, \\
  \end{split}
\end{equation}
or
\begin{equation}\label{Solution2}
  \begin{split}
    R_E+P_E&=T_E+S_E \\
    \alpha  &= \frac{2(\eta-1) P_E  + (1 - 2 \eta) T_E+ S_E}{(2 P_E - S_E - T_E) (S_E - T_E)} \\
    \beta    &=  \frac{2(1  -  \eta)P_E +(2 \eta- 1) S_E - T_E}{(2 P_E - S_E - T_E) (S_E - T_E)} \\
    \gamma   &= -\frac{\eta(S_E+T_E)}{2P_E-S_E-T_E}, \\
  \end{split}
\end{equation}
which means that only in the case that there are no errors ($\eta=0 \Leftrightarrow \epsilon=0,\xi=0$) as already proven in Appendix \ref{ap_2} or the case with $R_E+P_E=T_E+S_E$, TFT can enforce a linear payoff relationship.
Figure~\ref{figure3} shows the case with $R_E+P_E\neq T_E+S_E$.
When there are no errors (black dots), TFT can enforce a linear payoff relationship.
However, in the other cases (green and light green dots), the linear relationship collapses.

\section*{Acknowledgment}
This study was partly supported by HAYAO NAKAYAMA Foundation for Science \& Technology and Culture and JSPS KAKENHI Grant Number JP19K04903 (G.I.).


\end{document}